\documentstyle[12pt,epsfig]{article}

\begin{document}
\begin{titlepage}
\begin{center}
{\Large\bf On the Relation Between x-dependence  of the \\
[3mm] Higher Twist Contribution  to $F_3$ and $g_1^p - g_1^n$}
\end{center}
\vskip 2cm
\begin{center}
{\bf A. V. Sidorov}\\
{\it Bogoliubov Theoretical Laboratory\\
Joint Institute for Nuclear Research, 141980 Dubna, Russia }
%E-mail: sidorov@thsun1.jinr.ru}
\end{center}

\vskip 0.3cm
\begin{abstract}
\hskip -5mm We compare the higher twist (HT) contribution to the
unpolarized structure function $F_3$ with that one to the
nonsinglet combination $g_1^p - g_1^n$ of the polarized proton
and neutron structure functions using the assumption that the HT
contributions to the Gross-Llewellyn Smith and the Bjorken sum
rules are similar. We have found, that  the relation
$\frac{1}{3x}h^{xF_3}(x) \approx \frac{6}{g_A}h^{g_1^p -
g_1^n}(x)$ is valid for $x \geq 0.1$ and for $x \geq 0.2$ in the
case of LO and NLO QCD approximations, respectively.\\



\end{abstract}
\end{titlepage}

\setcounter{page}{1}

%\section{Introduction}
The structure functions in deep inelastic lepton nucleon
scattering are presently a subject of intensive experimental and
theoretical investigations. While the leading twist (LT) part of
the structure functions related with the parton distributions and
their $Q^2$-evolution is studied in detail in pQCD, the higher
twist corrections ($\sim1/Q^2$) are of a big interest and an
intensive study in the last years. The higher twist effects are
especially important in the case of polarized structure functions
because the most of the precise data (JLAB, HERMES, SLAC) are in
the region of $Q^2 \sim 1\; {\rm GeV}^2$.

In this note we consider the relation between the HT contribution
to the unpolarized structure function $F_3$ and $g^p_1-g^n_1$
which are pure non-singlets. As it was shown in the paper
\cite{Kodaira} the $Q^2$-evolutions of the $F_3$ and the
nonsinglet part of the  $g_1$ structure functions are identical up
to NLO order. Moreover the x-shapes of the $F_3$ and nonsinglet
part of $g_1$ are also similar \footnote{This property is
intensively used in  the phenomenological applications
\cite{Kotikov}.}. By analogy, one could suppose that the HT
contributions to $F_3$ and $g^p_1-g^n_1$ are similar too. Such an
assumption was recently used for the first moments of the HT
corrections in the Gross-Llewellyn Smith and Bjorken sum rules in
the infrared renormalons approach \cite{KataevGLSBjp}:
\begin{eqnarray}
GLS(Q^2)=\int^1_0 dx F_3 (x,Q^2) & = &
3(GLS_{pQCD}-\frac{\langle\langle O_1 \rangle\rangle}{Q^2})  \label{GLS}  \\
Bjp(Q^2)=\int^1_0 dx [g^{p}_1 (x,Q^2)-g^{n}_1 (x,Q^2)] & = &
\frac{g_A}{6}(Bjp_{pQCD}-\frac{\langle\langle O_2
\rangle\rangle}{Q^2})  \label{Bjp}
\end{eqnarray}
where
\begin{eqnarray}
\langle\langle O_1 \rangle\rangle & \approx &\langle\langle O_2
\rangle\rangle
 \label{HTGLSBJ}
\end{eqnarray}
Here $GLS_{pQCD}$ and $Bjp_{pQCD}$ are the leading twist
contribution to corresponding sum rules:
\begin{eqnarray}
GLS_{LO}=Bjp_{LO}& = & 1 \\
GLS_{NLO}=Bjp_{NLO} & = & 1-\alpha_S(Q^2)/\pi
 \label{SRlonlo}
\end{eqnarray}

In this note we are going to verify if the relation (3) between
the lowest moments of the HT contribution can be generalized for
the higher twists themselves, namely:
\begin{eqnarray}
\frac{1}{3x}h^{xF_3}(x)& \approx & \frac{6}{g_A}h^{g_1^p -
g_1^n}(x)
 \label{HTF3g1}
\end{eqnarray}
To test this relation we will use the values of HT obtained in
the QCD analyses of the corresponding structure functions in model
independent way. In the QCD analyses of DIS data when the higher
twist corrections are taken into account, the structure functions
are given by:
\begin{eqnarray}
xF_3 (x,Q^2) & = & xF_3(x, Q^2)_{\rm LT}
+ h^{xF_3}(x)/Q^2~ \\
g^p_1(x, Q^2) &  = & g^p_1(x, Q^2)_{\rm LT}  + h^{g_1^p}(x)/Q^2~ \\
g^n_1(x, Q^2) & = & g^n_1(x, Q^2)_{\rm LT}  + h^{g_1^n}(x)/Q^2~.
\label{LTHT}
\end{eqnarray}

\begin{figure}[t]
\centerline{ \epsfxsize=3.1in\epsfbox{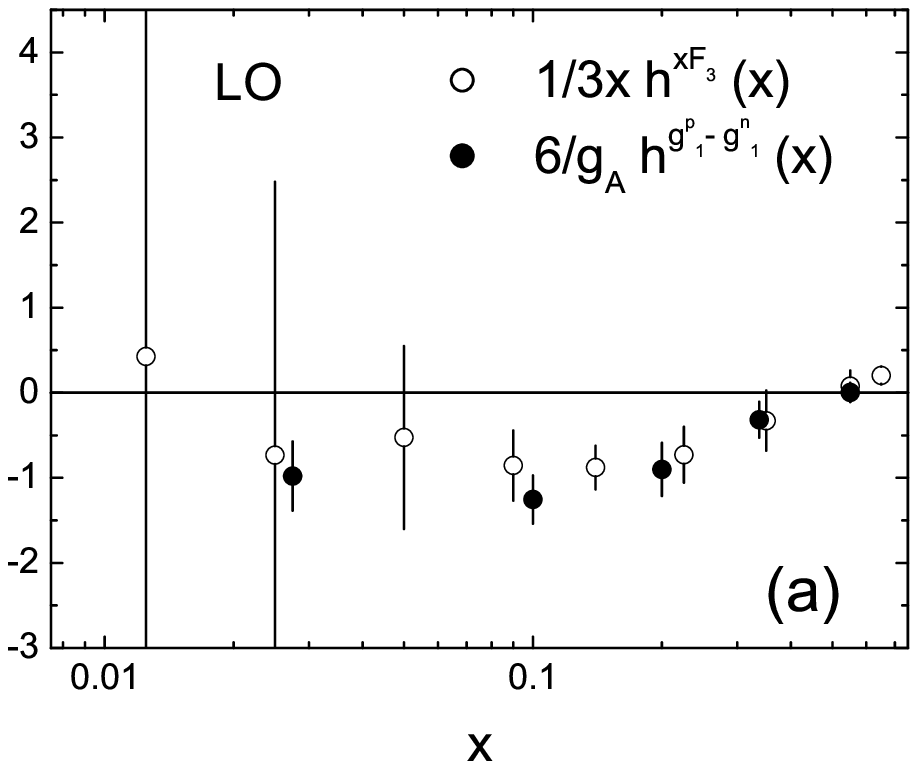}
\epsfxsize=3.1in\epsfbox{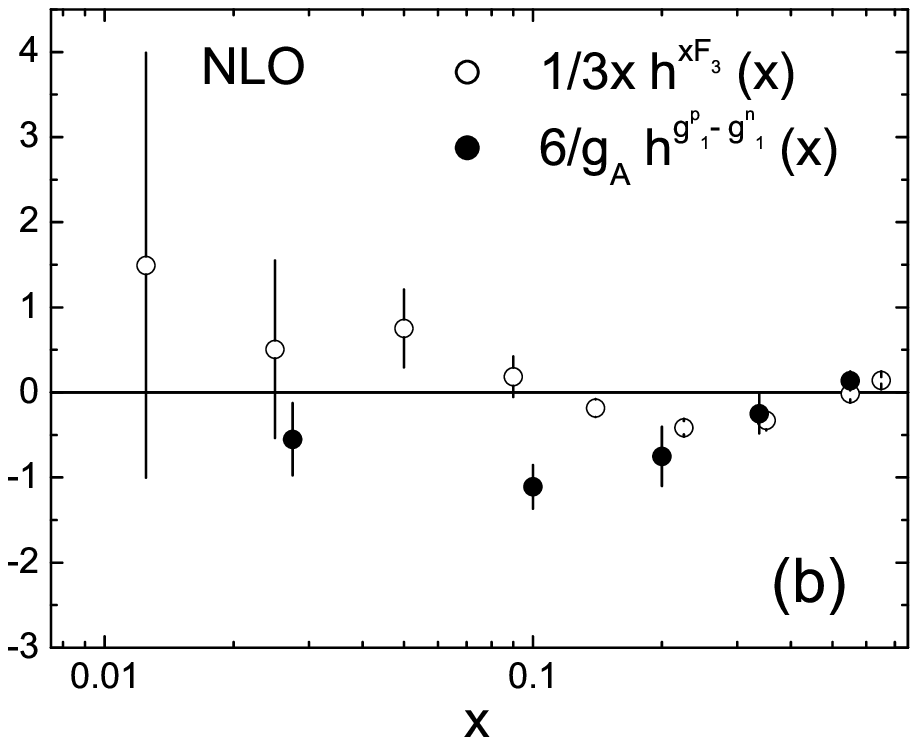} }\caption{Comparison of the LO
and NLO($\rm \overline{MS}$) results for $\frac{1}{3x}h^{xF_3}(x)$
based on the analysis of the CCFR data \cite{KPS,CCFRdata} - (open
cycles), and for $\frac{6}{g_A}h^{g_1^p - g_1^n}(x)$ based on  the
results of \cite{JHEP} - black cycles. \label{inter1}}
\end{figure}
\begin{figure}[thb]
\centerline{ \epsfxsize=3.1in\epsfbox{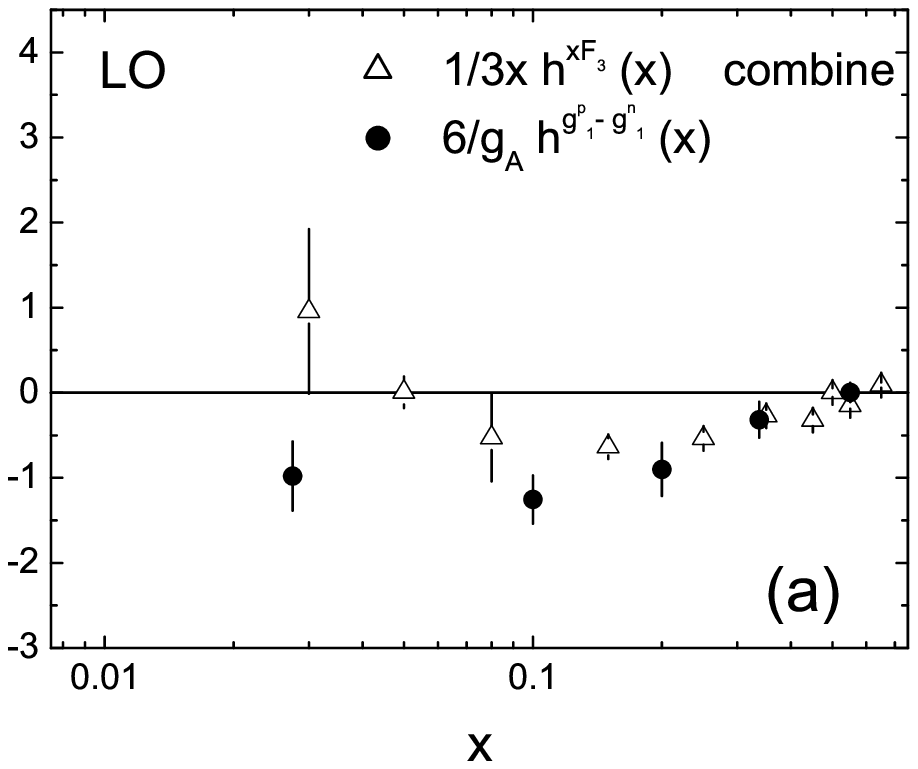}
\epsfxsize=3.1in\epsfbox{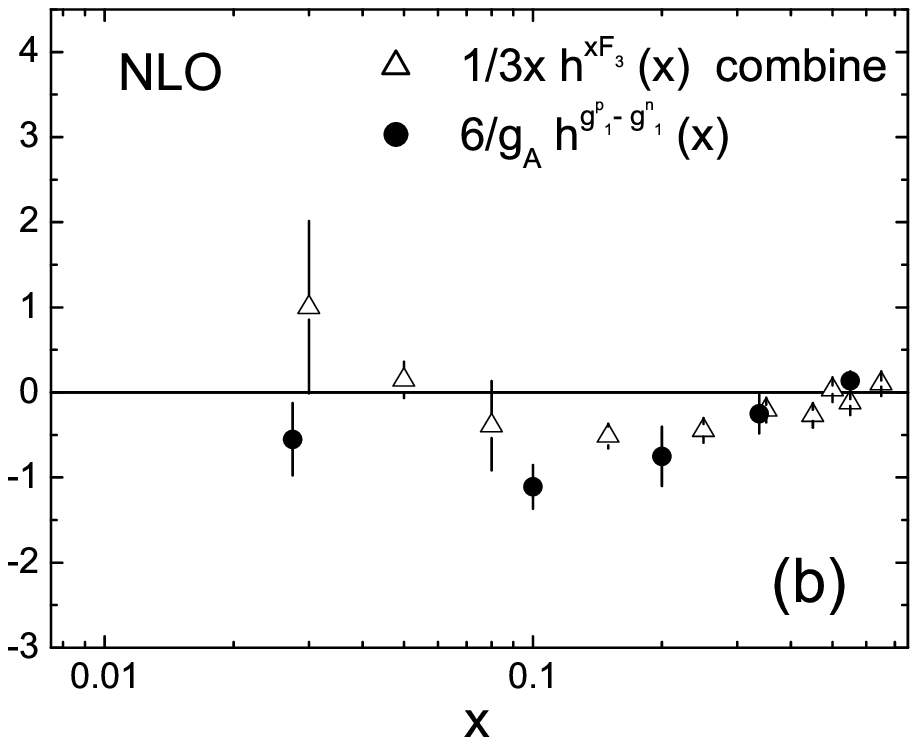} }
 \caption{
Comparison of  the  LO and NLO($\rm \overline{MS}$) results for
$\frac{1}{3x}h^{xF_3}(x)$ based on  the combined data analysis
\cite{combine,combinedata} - (open triangles), and for
$\frac{6}{g_A}h^{g_1^p - g_1^n}(x)$ based on  the results of
\cite{JHEP} - black cycles. \label{inter2}}
\end{figure}
\begin{figure}[th]
\centerline{ \epsfxsize=3.1in\epsfbox{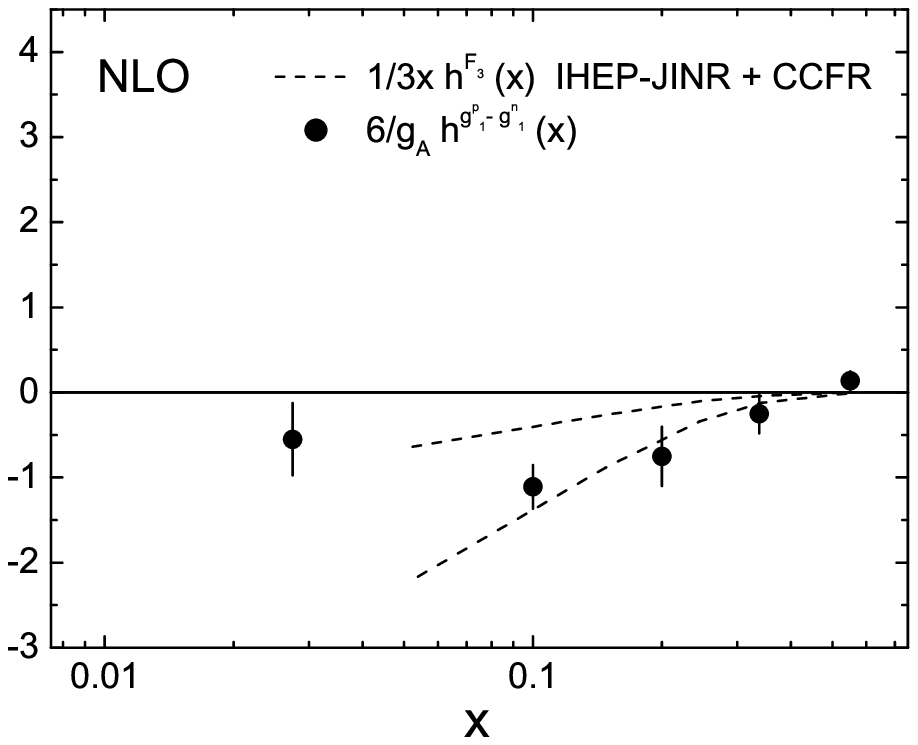} }
 \caption{
Comparison of  the NLO($\rm \overline{MS}$) results for
$\frac{1}{3x}h^{xF_3}(x)$ based on  the combined analysis of
IHEP-JINR \cite{Alekhin,IHEP-JINR} and CCFR data (desh lines
corresponds to upper and low limits of the infrared renormalon HT
contribution), and for $\frac{6}{g_A}h^{g_1^p - g_1^n}(x)$ based
on the results of \cite{JHEP} - black cycles. \label{inter2}}
\end{figure}
In (\ref{LTHT}) $h^{xF_3}(x), h^{g_1^p}(x)$ and $h^{g_1^n}(x) $
are the {\it dynamical} higher twists corrections to $xF_3$,
$g^p_1$ and $g_1^n$, which are related to multi-parton
correlations in the nucleon. They are non-perturbative effects and
can not be calculated without using models. The target mass
corrections, which are also corrections of inverse powers of
$Q^2$, are calculable {\cite{WW,TB} and effectively belong to the
leading twist term. A model independent determination of
$h^{xF_3}(x)$ was done in \cite{KPS}\footnote{See Table 12 in
\cite{KPS}} on the basis of the analysis of CCFR-NuTev
(anti-)neutrino deep--inelastic scattering data \cite{CCFRdata} at
$Q^2 \geq 5~GeV^2$  and in \cite{combine} using the combine set of
data \cite{combinedata} different from that of CCFR at $Q^2 \geq
0.5~GeV^2$. We consider also the results of \cite{Alekhin} where
the  infrared renormalon model approach for HT contribution was
applied in analysis of combine set of IHEP-JINR\cite{IHEP-JINR}
and CCFR-NuTev data. The values of $h^{g_1^p}(x)$ and
$h^{g_1^n}(x)$ in LO \footnote{It should be stressed, that LO
approach for QCD analysis of polarised structure function $g_1$ is
not reliable enough. See discussion in \cite{LSS2001,LSSHT}}} and
NLO($\rm \overline{MS}$) are given in \cite{JHEP}, where the
results of the analysis of the world data on polarized structure
function $g_1$ \cite{world} at $Q^2 \geq 1~GeV^2$, including the
precise JLab $g_1^n$ \cite{JLab} data, are presented. Using these
results and taking into account the coefficients in (\ref{GLS})
and (\ref{Bjp}) one could construct the l.h.s. and r.h.s. of Eq.
(\ref{HTF3g1}).

In Fig. 1, 2 and 3 we compare the results on HT in l.h.s. and
r.h.s. of Eq. (\ref{HTF3g1}). One can see (Fig. 1), that while in
the polarized case the values of HT change slightly from LO to NLO
approximation, in the unpolarized one the shape of $h^{xF_3}(x)$
depends on the order of pQCD used, especially for $ x\leq 0.1$.
 As
seen from Fig. 1, 2 and 3 the equality (\ref{HTF3g1}) is
approximately valid for $x \geq 0.1$ and $x \geq 0.2$ for the LO
and NLO approximations, respectively. It means that the higher
Mellin moments of the both parts of equation (\ref{HTF3g1}) should
approximately coincide:
\begin{eqnarray}
\int^1_0 dx ~x^N \frac{1}{3x}h^{xF_3}(x)&  \approx & \int^1_0 dx
~x^N \frac{6}{g_A}h^{g_1^p - g_1^n}(x), ~N - large.
 \label{momHTF3g1}
\end{eqnarray}

We would like to mention, that equality (3) is suggested in the
framework of the infrared renormalon approach, so the violation of
equality (4), which is shown in the Fig. 1b, Fig. 2a and 2b at $x
< 0.1$, could be due to the contribution of the dynamical twists
connected with the non-perturbative structure of the nucleon in
this $x$ region.

Finally, it should be noted, that there is additional sources of
uncertainties which should be taken into account in a more
detailed test of Eq. (\ref{HTF3g1}): the contribution of ${\cal
O}(1/Q^4)$); the separation of the twist-3 contribution in the
polarized case, which is effectively included in $h^{g_1}(x)$; the
$Q^2$ dependence of the functions $h(x)$, etc. \\ [4mm]

I am grateful for the hospitality of the Theory Group at CERN
where this work has been completed. This research was supported by
the RFBR grants N 05-02-17748 and N 05-01-00992. I am grateful for
D.B. Stamenov and C. Weiss for discussions.

\end{document}